\documentclass[prc,aps,twocolumn,showkeys,showpacs,superscriptaddress]{revtex4-2} 
\usepackage{graphicx}  
\usepackage{dcolumn}   
\usepackage{bm}        
\usepackage{amssymb}   
\usepackage{amsmath}    
\usepackage{xcolor}
\usepackage{makecell}
\usepackage{multirow}

\begin{document} 
 
\title{Rethinking Partial Widths: Unitary Mixing and the $\Delta(1232)$ Pole Residue}

\author{S.~Ceci}
\email{sasa.ceci@irb.hr}
\affiliation{Rudjer Bo\v{s}kovi\'{c} Institute, Bijeni\v{c}ka  54, HR-10000 Zagreb, Croatia}
\author{R.~Omerović} \affiliation{University of Tuzla, Urfeta Vejzagića 4, 75000 Tuzla, Bosnia and Herzegovina}
\author{H.~Osmanović} \affiliation{University of Tuzla, Urfeta Vejzagića 4, 75000 Tuzla, Bosnia and Herzegovina}
\author{M.~Uroić}
\affiliation{Rudjer Bo\v{s}kovi\'{c} Institute, Bijeni\v{c}ka  54, HR-10000 Zagreb, Croatia}
\author{B.~Zauner}
\affiliation{Institute for Medical Research and Occupational Health, Ksaverska 2, HR-10000 Zagreb, Croatia}

\begin{abstract}
The extracted $\pi N$ partial decay width of the $\Delta(1232)$ systematically exceeds its total width ($2|r|>\Gamma$). We demonstrate this anomaly is a natural consequence of S-matrix unitary mixing. Because exact multi-channel shadow poles are distant and model-dependent, we utilize a heuristic elastic model---treating the overlapping $\Delta(1600)$ as fully elastic---to isolate the core mechanism. We show that evaluating a perturbing S-matrix at a state's complex pole systematically inflates the residue magnitude. This proof of principle confirms complex residues reflect global amplitude topology rather than isolated intrinsic properties, challenging naive interpretations of branching fractions.
\end{abstract}


\maketitle

\section{Introduction}

In hadron physics, resonance properties are fundamentally defined by the complex poles ($W_p = M-i\,\Gamma/2$) and residues ($r = |r|e^{i\theta}$) of the scattering amplitude \cite{BreitWigner1936, DalitzMoorhouse1970}. While the pole position defines the resonance mass and total width ($\Gamma$), the residue magnitude $|r|$ is conventionally interpreted as the partial decay half-width \cite{PDG}. For a purely elastic state, this identification dictates that $2\,|r| = \Gamma$. However, this leads to a persistent physical puzzle for the $\Delta(1232)$. Leading partial-wave analyses \cite{Hoferichter2024,Ronchen2022,Svarc2014,Anisovich2012,Cutkosky1980} consistently extract residue magnitudes that violate this boundary, yielding $2\,|r|>\Gamma$. As detailed in Table \ref{tab:Delta1232}, the weighted average of this ratio across major modern models is approximately $104.9\pm1.7\%$, a statistically significant deviation that persists regardless of the specific analytic continuation method used.

\begin{table}[h]
    \centering
        \setlength{\tabcolsep}{4pt} 
\resizebox{\columnwidth}{!}
    {%
    \begin{tabular}{llll}
    \hline\hline
        $\mathbf{\Delta(1232)}$   &  $2\,|r|$  & $\Gamma$  & $2\,|r|/\Gamma$  \\
      &   (MeV) &  (MeV) &  (\%) \\
     \hline
PDG \cite{PDG}   & $100\pm 4$  & $100^{+4}_{-2}$  & $100^{+6}_{-4}$ \\
\hline
Hoferichter \cite{Hoferichter2024} & $102.6\pm1.8$ & $98.5\pm1.2$ & $104.2\pm2.2$\\
Roenchen \cite{Ronchen2022}    & $100\pm 2$ & $93\pm 1$ & $107.5\pm 2.4$ \\
\v Svarc \cite{Svarc2014}      & $100.0\pm 2.8$ & $98.0\pm 2.2$ & $102.0\pm 3.7$\\
Anisovich \cite{Anisovich2012} & $103.2\pm 1.2$ & $99\pm 2$ & $104.2\pm 2.4$ \\
Cutkosky \cite{Cutkosky1980}   & $106\pm 4$ & $100\pm 2$ & $106.0\pm 4.5$ \\
       \hline
Average        & $102.4\pm1.5$  & $96.4\pm2.9$  & $104.9\pm1.7$ \\
       \hline\hline
    \end{tabular}
    }
    \caption{
    The partial width puzzle: the $\pi N$ elastic partial decay width of the $\Delta(1232)$ (conventionally given by $2\,|r|$) is consistently extracted to be larger than the total decay width ($\Gamma$). The statistical average of the ratio across major models is $\bm{(104.9\pm1.7)\%}$.  
    }
    \label{tab:Delta1232}
\end{table}

This discrepancy is often dismissed as a minor background artifact. However, recent studies have demonstrated that the residue phase $\theta$ is not an isolated intrinsic property of the resonance, but is instead strongly dictated by external kinematic constraints, such as reaction thresholds and amplitude zeros \cite{Ceci26,Ceci26A}. In this Letter, we extend this paradigm to the residue magnitude, proposing that $|r|$ is similarly subject to systematic enhancement due to S-matrix unitarity.

By implementing a minimalist unitary framework, we show that the anomalous partial width of the $\Delta(1232)$ is a natural mathematical consequence of unitary mixing with higher-mass states in the same partial wave, specifically the $\Delta(1600)$. Our findings suggest that rather than being an isolated resonance parameter, the complex pole residue characterizes the global structure of the unitarized scattering amplitude. This underscores the necessity of rethinking how branching fractions and partial widths are interpreted in modern hadronic spectroscopy.

\section{Theoretical framework}

We begin with the Breit-Wigner formula \cite{BreitWigner1936}
\begin{equation}
    T_R=\frac{\Gamma_\mathrm{par}/2}{M-W-i\,\Gamma_\mathrm{tot}/2}
\end{equation}
where $W$ is the center-of-mass energy, and we conventionally define the branching fraction as $x=\Gamma_\mathrm{par}/\Gamma_\mathrm{tot}$.

For elastic resonances, such as the $\Delta(1232)$, \mbox{$\Gamma_\mathrm{par}=\Gamma_\mathrm{tot}$}, yielding the elastic Breit-Wigner formula 
\begin{equation}
    T_R=\frac{\Gamma/2}{M-W-i\,\Gamma/2},
\end{equation}
which by construction satisfies the elastic unitarity condition $|T_R|^2=\mathrm{Im} \, T_R$.

The full scattering amplitude $T$, containing left-hand cuts and contributions of other resonances, is modeled with the background term $T_B$. For an elastic resonance, following H\"{o}hler \cite{HohlerBible}, we write this form as
\begin{equation}
    T= T_R+T_B + 2i\,T_RT_B, \label{eq:HoehlerAddition}
\end{equation}
which satisfies S-matrix unitarity automatically if $T_B$ itself satisfies it. Alternatively, modern analyses often employ the $K$-matrix formalism to explicitly preserve unitarity when adding multiple resonant contributions:
\begin{equation}
    T = \frac{K}{1 - iK}, \quad \text{where} \quad K = K_1 + K_2 + \dots \label{eq:KmatrixForm}
\end{equation}

More general approaches typically model the width parameter $\Gamma$ giving it energy dependence that accounts for the threshold behavior, as well as the higher energy behavior. The unexpected result from Refs.~\cite{Ceci26, Ceci26A} shows us that near the pole we may absorb that energy dependence into a constant unitary background phase $\beta$. The amplitude for a single channel then becomes 
\begin{equation}
    T_1=\frac{x\,\Gamma/2\,e^{2i\beta}}{M-W-i\,\Gamma/2}+x\,e^{i\beta}\sin\beta, \label{Eq:ConstantBackgroundModel}
\end{equation}
where $\beta$ parameterizes the proximity of the effective amplitude zero to the threshold, and is directly related to the Breit-Wigner mass shift \cite{Lichtenberg1974, Manley1995}:
\begin{equation}
    \beta=-\arctan\frac{M_\mathrm{BW}-M}{\Gamma/2}.
\end{equation}
The corresponding background-adjusted $K$-matrix parameterization of this isolated resonance is:
\begin{equation}
    K_1 = \frac{\Gamma_\mathrm{BW}/2}{M_\mathrm{BW} - W} + \tan\beta, \label{Eq:KmatrixBackground}
\end{equation}
where mathematical equivalence at the pole demands $\Gamma_\mathrm{BW} = \Gamma/\cos^2\beta$. 

From Eq.~(\ref{Eq:ConstantBackgroundModel}) it is easy to directly read the residue magnitude as $|r|=x\,\Gamma/2$, and the residue phase as $\theta=2\beta$. For an isolated elastic $\Delta(1232)$ with $x=1$, the ratio $2\,|r|/\Gamma$ is strictly 100\%. 

To demonstrate why experimental extractions consistently overshoot this value, we introduce the contribution from the nearby $\Delta(1600)$ resonance within the same partial wave. Here, we face a practical topological challenge. The physical pole of the $\Delta(1600)$ resides on a distant, multi-channel Riemann sheet. Determining exact shadow-pole parameters on the primary elastic sheet typically requires full non-diagonal, multi-channel K-matrices. Implementing these would introduce numerous unknown matrix elements, diluting our analysis with arbitrary free parameters. 

Rather than chasing exact numerical precision by guessing these multi-channel couplings, we construct a heuristic thought experiment. We ask a foundational question: what happens to the extracted $\Delta(1232)$ parameters in an idealized scenario where the overlapping $\Delta(1600)$ state is completely elastic in the $\pi N$ channel? 

By artificially setting the perturbing state's branching fraction to $x=100\%$, we replace the distant multi-channel background with a maximal purely elastic contribution. The goal of this simplified toy model is not to perfectly reproduce the $\sim 105\%$ experimental average, but to stress-test the definition of the partial width itself. If the introduction of purely unitary overlap inflates the naive $2|r|/\Gamma$ ratio away from $100\%$, it provides a clear proof of principle that the complex pole residue cannot be equated to a physically bounded branching fraction.

The background phase $\beta_1$ fixes the effective zero of the isolated $\Delta(1232)$ amplitude at $W_0 = M_1 + \frac{\Gamma_1^2}{4(M_1 - M_\mathrm{BW1})} \approx 1096$~MeV, comfortably near the physical $\pi N$ threshold. Because we require the perturbing $\Delta(1600)$ amplitude to share this exact same threshold zero $W_0$, its effective background phase $\beta_2$ is strictly dictated, which in turn defines its corresponding Breit-Wigner parameters ($M_\mathrm{BW2}$, $\Gamma_\mathrm{BW2}$).

Because derived parameters in such formalisms are highly correlated, we express each analytically in terms of purely independent PDG inputs ($M_1, \Gamma_1, M_\mathrm{BW1}$ for the first state; $M_2, \Gamma_2$ for the second) and evaluate their limits using standard variance expansion \mbox{($\sigma_f^2 = \sum (\partial f/\partial x_i)^2 \sigma_{x_i}^2$)}. All initial, derived, and uncertainty-propagated parameters used in our unitary mixing models are summarized in Table \ref{tab:Input}.

\begin{table}[h!]
    \centering
    \setlength{\tabcolsep}{5pt} 
    \resizebox{\columnwidth}{!}
    {%
    \begin{tabular}{l c c c c}
    \hline\hline
    & \multicolumn{2}{c}{$\bm{\Delta(1232)}$} & \multicolumn{2}{c}{$\bm{\Delta(1600)}$} \\
    \textbf{Parameter} & PDG & Derived  & PDG  & Derived  \\ \hline 
    $M$ (MeV) & $\bm{1210 \pm 1}$ & $\bm{1210 \pm 2}$ & $\bm{1510 \pm 50}$ & $\bm{1510 \pm 50}$ \\
    $\Gamma$ (MeV) & $\bm{100 \pm 2}$ & $\bm{98 \pm 4}$ & $\bm{250 \pm 50}$ & $\bm{250 \pm 50}$ \\
    $M_\mathrm{BW}$ (MeV) & $\bm{1232 \pm 2}$ & $\bm{1232 \pm 2}$ & $1570 \pm 70^\ast$ & $\bm{1548 \pm 48}$ \\
    $\Gamma_\mathrm{BW}$ (MeV) & $\bm{117 \pm 3}$ & $\bm{119 \pm 4}$ & $250 \pm 50^\ast$ & $\bm{273 \pm 64}$ \\
    $\beta$ $(^\circ)$ & $-$ & $\bm{-23.7 \pm 2.2}$ & $-$ & $\bm{-16.8 \pm 3.7}$ \\
    $x$ (\%) & $99.4$ & $\bm{100}$ & $10-24$ & $\bm{100^\dagger}$ \\
    \hline\hline
    \end{tabular}
    }
    \caption{
    Input and derived parameters used for the unitary mixing models. Values in \textbf{boldface} are the ones explicitly used in our calculations, including rigorous error propagation limits. 
    For $\Delta(1232)$, the derived $\beta$ is fixed by the PDG pole and $M_\mathrm{BW}$. Converting the K-matrix (using PDG $\Gamma_\mathrm{BW} = 117\pm3$~MeV) back to the T-matrix yields a recovered pole. The small deviation in the recovered width ($98 \pm 4$~MeV vs.\ input $100 \pm 2$~MeV) illustrates the slight mathematical inconsistency in the nominal PDG parameter set (the exact theoretical BW width required to perfectly match the pole is $\Gamma_\mathrm{BW} \approx 119 \pm 4$~MeV).
    ($^\ast$) For $\Delta(1600)$, the nominal PDG BW parameters are shown for reference only and are not used in our calculations. 
    ($^\dagger$) To approximate the $\Delta(1600)$ shadow-pole contribution purely for our proof of principle, we artificially set $x=100\%$ (exactly) and tune its phase ($\beta = -16.8^\circ$) so the amplitude zero perfectly coincides with the threshold zero of the $\Delta(1232)$. Because we calculate its $M_\mathrm{BW}$ and $\Gamma_\mathrm{BW}$ consistently from this $\beta$ and the PDG pole, the recovered K-matrix pole exactly tracks the input with its corresponding uncertainties.
    }
    \label{tab:Input}
\end{table}

\section{Results}

To demonstrate how unitarity inherently inflates the residue magnitude, we implement a step-by-step analytical addition of the resonant states. Because exact numerical precision is not the goal of a minimalist toy model, we focus on the direction and scale of the effect. We evaluate the system using two independent unitary formalisms: the H\"{o}hler T-matrix addition (where $S = S_1 S_2 \dots$) and the modern K-matrix framework (Eq.~(\ref{eq:KmatrixForm})).

\subsection{Unitary Inflation of the Residue}

If the $\Delta(1232)$ were the only resonance in this partial wave, the extracted residue magnitude $|r|$ would exactly equal $x\,\Gamma/2$. In that isolated elastic limit, the partial decay width ratio is strictly $100\%$. 

However, introducing the generic $\Delta(1600)$ perturbation modifies the amplitude. In the T-matrix framework, combining the full background-adjusted amplitudes from Eq.~(\ref{Eq:ConstantBackgroundModel}), the original residue ($r_1$) is multiplied by an interference factor generated by the tail of the second resonance evaluated at the first pole's position ($W_1 = M_1 - i\Gamma_1/2$). This factor is exactly the S-matrix of the second resonance, $S_2$:
\begin{equation}
    r_1^\mathrm{new} = r_1 S_2(W_1). \label{eq:ResidueStretching}
\end{equation}

The fundamental mechanism driving this enhancement is the analytic continuation of S-matrix unitarity. While resonant S-matrix contributions are strictly unitary on the real energy axis ($|S(W_\mathrm{real})| = 1$), they are not constrained to be unitary deep in the complex plane. Consequently, the modulus of the perturbing S-matrix evaluated exactly at the complex pole position of the first resonance is generally strictly greater than unity ($|S_2(W_1)| > 1$). Because $S_2(W_1)$ is a complex number evaluated in the presence of overlapping tails, its direct interference mathematically stretches the residue magnitude $|r_1^\mathrm{new}|$, causing the partial decay width ratio $2|r|/\Gamma$ to jump from $100\%$ to $\bm{112 \pm 3\%}$. The propagated uncertainty reflects the large $\pm 50$~MeV uncertainties of the PDG inputs for the perturbing state. 

To ensure this enhancement is not an artifact of H\"{o}hler's product approximation, we test the exact same parameters using K-matrix unitarity. It is important to mathematically note that a simple additive K-matrix formulation ($K = K_1 + K_2$) is not algebraically identical to the multiplicative S-matrix ($S = S_1 S_2$) implicitly generated by H\"{o}hler's addition. The additive K-matrix inherently generates cross-terms in the denominator that slightly shift the complex pole position compared to the T-matrix framework. 

Despite this structural difference in the analytic continuation, the newly extracted partial width ratio in the K-matrix framework is practically identical ($\bm{112 \pm 3\%}$). The fact that two phenomenologically distinct unitary formalisms converge to the same anomalous enhancement confirms that this inflation is not a quirk of a specific mathematical scheme, but a fundamental, unavoidable topological property of overlapping poles in the unitarized S-matrix.

\subsection{Stability and the Third Resonance}

To evaluate whether this inflation is a stable feature, we extend our analysis by including the next higher-lying state in this partial wave, the $\Delta(1920)$. To ensure complete transparency and consistency with our methodology for the $\Delta(1600)$, we ground this perturbation in the nominal PDG estimates: $M_3 = 1900 \pm 50$~MeV, $\Gamma_3 = 300 \pm 100$~MeV, and an inelastic branching fraction of $x \approx 12.5\%$. 

For our analytical proof of principle, we again set the branching fraction artificially to $x=100\%$ and constrain the amplitude to share the exact same physical threshold zero ($W_0 \approx 1096.4$~MeV) as the primary state. This rigorous constraint analytically dictates the derived parameters for this third state: its background phase is fixed to $\beta_3 \approx -10.6^\circ$, which consequently defines its Breit-Wigner mass ($M_\mathrm{BW3} \approx 1928$~MeV) and width ($\Gamma_\mathrm{BW3} \approx 310$~MeV).

In the T-matrix framework, this multi-resonance addition is handled recursively. We take the already unitarized two-state amplitude $T_{12} = T_1 + T_2 + 2i\,T_1 T_2$ and combine it with the third amplitude $T_3$ using the exact same H\"{o}hler prescription:
\begin{equation}
    T_{123} = T_{12} + T_3 + 2i\,T_{12} T_3.
\end{equation}
When evaluating the residue of the $\Delta(1232)$ at its pole position ($W_1$), this recursive addition translates into a beautifully simple sequential multiplication. The new residue picks up an additional interference factor directly from the third state:
\begin{align}
    r_1^\mathrm{new} &= r_1 \left[ 1 + 2i\,T_2(W_1) \right] \left[ 1 + 2i\,T_3(W_1) \right] \nonumber \\
    &= r_1 S_2(W_1) S_3(W_1).
\end{align}
This demonstrates the underlying multiplicative nature of S-matrix poles. The inclusion of the $\Delta(1920)$ generates this additional complex interference factor ($S_3$), which consistently drives the $\Delta(1232)$ residue magnitude even higher.

For theoretical completeness, we also implement this three-state mixing in the K-matrix formalism by simply extending the background-adjusted sum:
\begin{equation}
    K(W) = K_1(W) + K_2(W) + K_3(W).
\end{equation}
As summarized in Table \ref{tab:Results}, both formalisms again converge to an almost identical cumulative enhancement (reaching $\sim 115\%$). 

This confirms that the partial width anomaly is a robust, cumulative effect of unitary mixing: any additional resonant state overlapping in the same partial wave systematically contributes to the inflation of the naive partial decay width.

\subsection{The Crucial Role of Background Phases}

A final critical insight comes from analyzing the background phases ($\beta$). To confirm that unitary mixing is the primary driver of the anomaly, we performed a theoretical stress test by mathematically removing the background phases ($\beta_i = 0$) and mixing only the bare Breit-Wigner functions. 

Physically, this breaks the correct threshold behavior. Mathematically, it isolates the unmitigated power of direct pole interference. Without the background phases, the K-matrix poles shift drastically (for instance, the $\Delta(1232)$ pole mass artificially jumps by over $40$~MeV), whereas the T-matrix framework rigidly locks the poles but still generates extreme residue stretching. Despite these wildly different structural responses, the branching fraction ratio across both models stubbornly increases to roughly $\bm{117\%}$. The residue phase, however, experiences a massive, non-physical rotation (jumping to approximately $+8^\circ$). 

This reveals a critical physical distinction in how unitary mixing affects resonance parameters. As established in our previous work, the physical value of the residue phase is driven by a delicate interplay between the pole positions, the Breit-Wigner mass, and the background constraints. When the background is removed, the phase fluctuates wildly across different unitary additions. 

In stark contrast, the magnitude of the residue—and thus the anomalous partial width—remains remarkably robust. Regardless of whether the background is present or absent, and despite the resulting massive shifts in pole positions, the branching fraction ratio stubbornly persists well above the $100\%$ limit. 

Therefore, while the background acts as a vital unitary regulator necessary to tame the phase to physical values and anchor the unitarized poles at their correct physical locations, the fundamental anomaly of $2|r| > \Gamma$ is an unavoidable, structural consequence of S-matrix pole overlap.

\begin{table}[h!]
    \centering
    \setlength{\tabcolsep}{8pt} 
    \resizebox{\columnwidth}{!}
    {%
    \begin{tabular}{lcc}
    \hline\hline
    \textbf{Model Scenario} & $\bm{\theta}$ & $\bm{2\,|r|/\Gamma}$ \\
    for the $\Delta(1232)$ state & $\mathbf{(^\circ)}$ & \textbf{(\%)} \\ \hline 
    \color{gray} Isolated Resonance & \color{gray} $-47$ & \color{gray} $100$ \\
    Two-State Mixing (T-matrix) & $-37 \pm 5$ & $112 \pm 3$ \\
    Two-State Mixing (K-matrix) & $-46 \pm 5$ & $112 \pm 3$ \\
    Three-State Mixing (T-matrix) & $-30 \pm 8$ & $115 \pm 5$ \\
    Three-State Mixing (K-matrix) & $-39 \pm 8$ & $115 \pm 5$ \\
    Bare BW Mixing (No $\beta$ phases) & $+8$ & $>117$ \\ \hline
    PDG Extracted \cite{PDG} & $-46_{-2}^{+1}$ & $105 \pm 2$ \\
    \hline\hline
    \end{tabular}
    }
    \caption{
    Comparison of the analytical results for the $\Delta(1232)$ residue phase and partial width ratio. Unitary mixing systematically inflates the ratio above the physically isolated limit (gray). Adding a third resonance ($\Delta(1920)$) proves the effect is cumulative across both T-matrix and K-matrix frameworks. Removing the background phases ($\beta$) shows that threshold constraints actively regulate the otherwise extreme unitary inflation. All uncertainties reflect the rigorous propagation of standard PDG errors.
    }
    \label{tab:Results}
\end{table}

\section{Discussion}

By deliberately scaling the $\Delta(1600)$ elasticity to 100\%, our heuristic model naturally acts as a magnifying glass for the unitary mixing effect. Consequently, our calculated enhancement ($\sim 112\%$) is expectedly larger than the statistical experimental average ($\sim 105\%$). However, the qualitative behavior is unequivocal: unitarity naturally drives the residue magnitude in the exact direction of the observed anomaly. We anticipate that incorporating further physical complexities—such as the exact multi-sheet dynamics and the explicit inclusion of inelastic channel openings ($\pi \Delta$, $\rho N$)—would naturally regulate and absorb the remaining numerical difference. The fact that such a simplified proof of principle captures the correct sign and a massive portion of the magnitude of the $\Delta(1232)$ puzzle strongly supports our central thesis.

The fundamental issue lies in the assumption that extracting a pole residue equates to isolating a pure resonance property. Even if complex pole positions represent true intrinsic properties of a state, their corresponding residues depend on all other resonances and thresholds sharing the same quantum numbers. Paradoxically, the very principle of S-matrix unitarity—which theoretically exists to ensure that physical branching fractions cannot exceed 100\%—is exactly what inflates the extracted pole residue beyond this physical limit when multiple states overlap.

Therefore, there is nothing inherently mathematically wrong with the extraction of complex pole residues; the problem arises only when one attempts to map them directly to intrinsic physical properties, such as a localized partial decay width. This misinterpretation is particularly troublesome in meson spectroscopy, where the magnitude of the residue is frequently utilized to define the fundamental coupling of a resonance to a specific decay channel \cite{PDG}. As our model demonstrates, that coupling will inevitably be "contaminated" by other states in the same partial wave. 

Physically, no state can decay into a given channel with a probability exceeding 100\%. A recent theoretical effort by Heuser {\it et al.} \cite{Heuser2024} proposes a rigorous new recipe to relate pole information to physically consistent branching fractions. That effort is highly welcome. However, the very necessity for a sophisticated mathematical approach designed to ensure that partial widths sum correctly further underscores our central argument: the direct, naive interpretation of the complex pole residue as an isolated physical property is untenable. The complex residue is, fundamentally, a global feature of the unitarized scattering amplitude.

\section{Conclusions}

In conclusion, this work provides compelling analytical evidence that the complex pole residue characterizes the global topological structure of the scattering amplitude rather than acting as an isolated, intrinsic property of the resonance itself. By implementing a minimalist, fully elastic unitary model---which conceptually maps the unknown multi-channel shadow poles onto the primary physical sheet---we have isolated the fundamental consequences of S-matrix mixing from model-dependent background noise.

Through two independent phenomenological frameworks (H\"{o}hler's T-matrix addition and K-matrix parameterization), we have demonstrated that the statistically significant extraction of $2\,|r|>\Gamma$ for the $\Delta(1232)$ is not a background artifact, but an unavoidable mathematical necessity of overlapping resonances. The core driver is the analytic continuation of S-matrix unitarity: while resonant contributions are strictly unitary on the real energy axis, they are not constrained to be unitary deep in the complex plane. Consequently, evaluating one state's overlapping amplitude at the complex pole position of another fundamentally distorts and systematically inflates the extracted residue magnitude.

Furthermore, we have shown that background phases---driven by physical threshold constraints---act as vital unitary regulators. They actively suppress what would otherwise be wild, non-physical interferences generated by this complex mixing, taming the anomalous residue enhancement down toward the experimentally observed scale of $\sim 105\%$.

These findings strongly suggest that the direct, naive identification of the complex residue magnitude with a localized, intrinsic partial decay width is physically problematic across hadronic spectroscopy. Because a pole residue is inherently modified by the tails of overlapping states in the same partial wave, we recommend that future phenomenological extractions and parameter evaluations, including those by the Review of Particle Physics, explicitly account for this intrinsic unitary inflation when interpreting branching fractions in multi-resonance systems.

\end{document}